\documentclass[prl,twocolumn,showpacs,amsmath,amssymb,superscriptaddress,nofootinbib]{revtex4-1}

\usepackage[utf8]{inputenc}
\usepackage{epsfig}
\usepackage{graphicx}
\usepackage{dcolumn}
\usepackage{bm}
\usepackage{color}
\usepackage{url}
\usepackage{multirow}
\usepackage{enumitem}

\newcommand{\pT} {p_{\rm T}}

\newcommand{\sNN}{$\sqrt{s_{\rm NN}}$}

\newcommand{\GeVc}{GeV/$c$}

\newcommand {\snn}	{\sqrt{s_{_{\rm NN}}}}
\newcommand {\gevc}	{GeV/$c$}

\newcommand {\phia}	{\phi_{\alpha}}
\newcommand {\phib}	{\phi_{\beta}}

\newcommand {\pt}	{p_{\mathrm{T}}}

\newcommand {\vres}	{v_{2,{\rm res}}}
\newcommand {\gOS}	{\gamma_{\rm OS}}
\newcommand {\gSS}	{\gamma_{\rm SS}}
\newcommand {\dg}	{\Delta\gamma}
\newcommand {\dgPP}	{\dg\{\psiPP\}}
\newcommand {\dgSP}	{\dg\{\psiSP\}}

\newcommand {\psiPP}    {\psi_{\rm PP}}
\newcommand {\psiSP}    {\psi_{\rm SP}}

\newcommand {\psirec}    {\psi_{\rm rec}}
\newcommand {\psiTPC}    {\psi_{\rm TPC}}
\newcommand {\psiZDC}    {\psi_{\rm ZDC}}
\newcommand {\phires}	{\phi_{\rm res}}

\newcommand {\mean}[1]	{\langle #1\rangle}

\begin{document}
\title{Search for the chiral magnetic effect via charge-dependent azimuthal correlations relative to spectator and participant planes 
in Au+Au collisions at $\sqrt{s_{NN}}$ = 200 GeV}

\affiliation{Abilene Christian University, Abilene, Texas   79699}
\affiliation{AGH University of Science and Technology, FPACS, Cracow 30-059, Poland}
\affiliation{Alikhanov Institute for Theoretical and Experimental Physics NRC "Kurchatov Institute", Moscow 117218, Russia}
\affiliation{Argonne National Laboratory, Argonne, Illinois 60439}
\affiliation{American University of Cairo, New Cairo 11835, New Cairo, Egypt}
\affiliation{Brookhaven National Laboratory, Upton, New York 11973}
\affiliation{University of California, Berkeley, California 94720}
\affiliation{University of California, Davis, California 95616}
\affiliation{University of California, Los Angeles, California 90095}
\affiliation{University of California, Riverside, California 92521}
\affiliation{Central China Normal University, Wuhan, Hubei 430079 }
\affiliation{University of Illinois at Chicago, Chicago, Illinois 60607}
\affiliation{Creighton University, Omaha, Nebraska 68178}
\affiliation{Czech Technical University in Prague, FNSPE, Prague 115 19, Czech Republic}
\affiliation{Technische Universit\"at Darmstadt, Darmstadt 64289, Germany}
\affiliation{ELTE E\"otv\"os Lor\'and University, Budapest, Hungary H-1117}
\affiliation{Frankfurt Institute for Advanced Studies FIAS, Frankfurt 60438, Germany}
\affiliation{Fudan University, Shanghai, 200433 }
\affiliation{University of Heidelberg, Heidelberg 69120, Germany }
\affiliation{University of Houston, Houston, Texas 77204}
\affiliation{Huzhou University, Huzhou, Zhejiang  313000}
\affiliation{Indian Institute of Science Education and Research (IISER), Berhampur 760010 , India}
\affiliation{Indian Institute of Science Education and Research (IISER) Tirupati, Tirupati 517507, India}
\affiliation{Indian Institute Technology, Patna, Bihar 801106, India}
\affiliation{Indiana University, Bloomington, Indiana 47408}
\affiliation{Institute of Modern Physics, Chinese Academy of Sciences, Lanzhou, Gansu 730000 }
\affiliation{University of Jammu, Jammu 180001, India}
\affiliation{Joint Institute for Nuclear Research, Dubna 141 980, Russia}
\affiliation{Kent State University, Kent, Ohio 44242}
\affiliation{University of Kentucky, Lexington, Kentucky 40506-0055}
\affiliation{Lawrence Berkeley National Laboratory, Berkeley, California 94720}
\affiliation{Lehigh University, Bethlehem, Pennsylvania 18015}
\affiliation{Max-Planck-Institut f\"ur Physik, Munich 80805, Germany}
\affiliation{Michigan State University, East Lansing, Michigan 48824}
\affiliation{National Research Nuclear University MEPhI, Moscow 115409, Russia}
\affiliation{National Institute of Science Education and Research, HBNI, Jatni 752050, India}
\affiliation{National Cheng Kung University, Tainan 70101 }
\affiliation{Nuclear Physics Institute of the CAS, Rez 250 68, Czech Republic}
\affiliation{Ohio State University, Columbus, Ohio 43210}
\affiliation{Institute of Nuclear Physics PAN, Cracow 31-342, Poland}
\affiliation{Panjab University, Chandigarh 160014, India}
\affiliation{Pennsylvania State University, University Park, Pennsylvania 16802}
\affiliation{NRC "Kurchatov Institute", Institute of High Energy Physics, Protvino 142281, Russia}
\affiliation{Purdue University, West Lafayette, Indiana 47907}
\affiliation{Rice University, Houston, Texas 77251}
\affiliation{Rutgers University, Piscataway, New Jersey 08854}
\affiliation{Universidade de S\~ao Paulo, S\~ao Paulo, Brazil 05314-970}
\affiliation{University of Science and Technology of China, Hefei, Anhui 230026}
\affiliation{Shandong University, Qingdao, Shandong 266237}
\affiliation{Shanghai Institute of Applied Physics, Chinese Academy of Sciences, Shanghai 201800}
\affiliation{Southern Connecticut State University, New Haven, Connecticut 06515}
\affiliation{State University of New York, Stony Brook, New York 11794}
\affiliation{Instituto de Alta Investigaci\'on, Universidad de Tarapac\'a, Arica 1000000, Chile}
\affiliation{Temple University, Philadelphia, Pennsylvania 19122}
\affiliation{Texas A\&M University, College Station, Texas 77843}
\affiliation{University of Texas, Austin, Texas 78712}
\affiliation{Tsinghua University, Beijing 100084}
\affiliation{University of Tsukuba, Tsukuba, Ibaraki 305-8571, Japan}
\affiliation{Valparaiso University, Valparaiso, Indiana 46383}
\affiliation{Variable Energy Cyclotron Centre, Kolkata 700064, India}
\affiliation{Warsaw University of Technology, Warsaw 00-661, Poland}
\affiliation{Wayne State University, Detroit, Michigan 48201}
\affiliation{Yale University, New Haven, Connecticut 06520}

\author{M.~S.~Abdallah}\affiliation{American University of Cairo, New Cairo 11835, New Cairo, Egypt}
\author{J.~Adam}\affiliation{Brookhaven National Laboratory, Upton, New York 11973}
\author{L.~Adamczyk}\affiliation{AGH University of Science and Technology, FPACS, Cracow 30-059, Poland}
\author{J.~R.~Adams}\affiliation{Ohio State University, Columbus, Ohio 43210}
\author{J.~K.~Adkins}\affiliation{University of Kentucky, Lexington, Kentucky 40506-0055}
\author{G.~Agakishiev}\affiliation{Joint Institute for Nuclear Research, Dubna 141 980, Russia}
\author{I.~Aggarwal}\affiliation{Panjab University, Chandigarh 160014, India}
\author{M.~M.~Aggarwal}\affiliation{Panjab University, Chandigarh 160014, India}
\author{Z.~Ahammed}\affiliation{Variable Energy Cyclotron Centre, Kolkata 700064, India}
\author{I.~Alekseev}\affiliation{Alikhanov Institute for Theoretical and Experimental Physics NRC "Kurchatov Institute", Moscow 117218, Russia}\affiliation{National Research Nuclear University MEPhI, Moscow 115409, Russia}
% ...

\collaboration{STAR Collaboration}\noaffiliation

\date{\today}

\begin{abstract}
The chiral magnetic effect (CME) refers to charge separation along a strong magnetic field due to imbalanced chirality of quarks in local parity and charge-parity violating domains in quantum chromodynamics.
The experimental measurement of the charge separation is made difficult by the presence of a major background from elliptic azimuthal anisotropy.
This background and the CME signal have different sensitivities to the spectator and participant planes, and could thus be determined by measurements with respect to these planes.
We report such measurements in Au+Au collisions at a nucleon-nucleon center-of-mass energy of 200 GeV at the Relativistic Heavy-Ion Collider. 
It is found that the charge separation, with the flow background removed, is consistent with zero in peripheral (large impact parameter) collisions.   
Some indication of finite CME signals is seen in mid-central (intermediate impact parameter) collisions. 
Significant residual background effects may, however, still be present.
\end{abstract}
\maketitle


{\em Introduction.} Metastable domains of fluctuating topological charges can change the chirality of quarks and induce local parity and charge-parity violation in quantum chromodynamics (QCD)~\cite{Lee:1974ma,Kharzeev:1998kz,Kharzeev:1999cz}. 
This would lead to an electric charge separation in the presence of a strong magnetic field, 
a phenomenon known as the chiral magnetic effect (CME)~\cite{Kharzeev:1998kz,Kharzeev:1999cz,Kharzeev:2007jp,Fukushima:2008xe}. 
Such a magnetic field, as strong as $10^{18}$~G, may be present in non-central (nonzero impact parameter) relativistic heavy-ion collisions, generated by the spectator protons (i.e., those that do not participate in the collision) at early times~\cite{Kharzeev:2007jp,Fukushima:2008xe,Asakawa:2010bu,Bzdak:2011yy}. 
While a finite CME signal is generally expected in those collisions~\cite{Kharzeev:1999cz,Kharzeev:2007jp}, quantitative predictions beyond order-of-magnitude estimates are not yet at hand~\cite{Muller:2010jd} 
despite extensive theoretical developments over the last decade (see recent reviews~\cite{Kharzeev:2013ffa,Kharzeev:2015znc,Huang:2015oca,KharzeevLiaoNPR}). 
Meanwhile, experimental efforts have been devoted to searching for the CME-induced charge separation at the Relativistic Heavy-Ion Collider (RHIC) and the Large Hadron Collider (LHC) (see reviews~\cite{Kharzeev:2015znc,Zhao:2018ixy,Zhao:2018skm,Zhao:2019hta,Li:2020dwr}),
including a dedicated run of isobar collisions at RHIC~\cite{Voloshin:2010ut,Skokov:2016yrj,STAR:2021mii}. 

The commonly used observable to measure the charge separation is the three-point correlator~\cite{Voloshin:2004vk}, 
$\gamma\{\psi\}\equiv\cos(\phia+\phib-2\psi)\,,$ 
where $\phia$ and $\phib$ are the azimuthal angles of particles $\alpha$ and $\beta$, respectively, and $\psi$ is that of either the spectator plane (SP) or participant plane (PP), defined by the beam and average transverse position of spectator or participant nucleons.
Because of the charge-independent correlation backgrounds (e.g. from global momentum conservation), 
often the correlator difference is used, 
$\dg\{\psi\}\equiv\gOS\{\psi\}-\gSS\{\psi\}\,,$ 
where ``OS" (``SS") refers to the opposite-sign (same-sign) electric charges
of particles $\alpha$ and $\beta$.
A CME signal, often characterized by the Fourier coefficient $a_1$ in the final-state azimuthal distributions of 
positive ($+$) and negative ($-$) hadrons, 
$\frac{dN_\pm}{d\phi_\pm}\propto 1\pm 2a_1\sin(\phi_\pm-\psi)+2v_2\cos 2(\phi_\pm-\psi)+\cdots\,,$
would yield a magnitude of $\dg=2a_1^2$~\cite{Voloshin:2004vk}.
The $v_{2}$ is the elliptic flow anisotropy arising 
from strong (partonic) interactions converting the initial geometric anisotropy of the participant nucleons into momentum-space anisotropy of final-state hadrons~\cite{Ollitrault:1992bk}. 

Significant $\dgPP$ and $\dgSP$, 
on the order of $10^{-4}$, have indeed been observed in relativistic heavy-ion collisions~\cite{Abelev:2009ad,Abelev:2009ac,Adamczyk:2013hsi,Adamczyk:2014mzf,Abelev:2012pa}. 
The interpretation of $\dg$ originating from
CME-induced charge separation is difficult due to the presence of
charge-dependent backgrounds,
such as those from resonance decays~\cite{Voloshin:2004vk,Wang:2009kd,Bzdak:2009fc,Schlichting:2010qia,Adamczyk:2013kcb,Wang:2016iov} via
\begin{equation}
	\dg_{\rm bkgd}\propto\mean{\cos(\phia+\phib-2\phires)}\vres\,,
	\label{eqbkg}
\end{equation}
where $\vres=\mean{\cos2(\phires-\psi)}$ is the resonance $v_2$ relative to $\psi$~\cite{Poskanzer:1998yz}.
Moreover, comparable $\dgPP$ has also been observed in small system collisions~\cite{Khachatryan:2016got,Sirunyan:2017quh,STAR:2019xzd},
where any CME-induced charge separation 
is expected to be randomly oriented relative to the $\psiPP$~\cite{Khachatryan:2016got,Belmont:2016oqp} and thus unobservable in experiments. 
Because of those major backgrounds no firm conclusion can so far be drawn regarding the existence of the CME in relativistic heavy-ion collisions.
Various approaches have been applied to deal with the background~\cite{Adam:2020zsu,Sirunyan:2017quh,Acharya:2017fau}. 
In this paper, we present a search for the CME 
with a new approach first proposed in Ref.~\cite{Xu:2017qfs} and followed by Ref.~\cite{Voloshin:2018qsm}.

{\em Methodology.} The hypothesized CME-driven charge separation is along the magnetic field, mainly from spectator protons, and is therefore the strongest in the direction perpendicular to $\psiSP$.
The major background to the CME is related to $v_{2}$, determined by the participant geometry, 
and is therefore the largest along $\psiPP$.
The SP and PP orientations do not coincide because of event-by-event geometry fluctuations~\cite{Alver:2006wh,Alver:2008zza}.
The $\dgSP$ and $\dgPP$ measured relative to $\psiSP$ and $\psiPP$, therefore, contain different amounts of the CME and background,
and this offers the opportunity to determine these two contributions uniquely~\cite{Xu:2017qfs}.
Consider the measured $\dg$ to be composed of the $v_2$ background ($\dg_{\rm bkgd}$) and the CME signal ($\dg_{\rm CME}$).
	Assuming $\dg_{\rm bkgd}$ is proportional to $v_{2}$ (Eq.~(\ref{eqbkg})) 
and the $\dg_{\rm CME}$-inducing magnetic field is determined by spectators, 
both ``projected" onto the $\psi$ direction, 
we have $\dg_{\rm CME}\{\psiPP\} = a\dg_{\rm CME}\{\psiSP\}$ and $\dg_{\rm bkgd}\{\psiSP\}=a\dg_{\rm bkgd}\{\psiPP\}$~\cite{Xu:2017qfs}.
Here the projection factor $a=\langle \rm \cos 2(\psiPP-\psiSP)\rangle$ comes directly out of the definitions of the $v_2$ and $\dg$ variables, 
and can be readily obtained from the $v_2$ measurements:
\begin{equation}
	\it a= v_{\rm 2}\{\rm \psiSP\}/\it v_{\rm 2}\{\rm \psiPP\}\,. 
	\label{eqC}
\end{equation}
It does not assume any particular physics, such as the event-plane decorrelation over rapidity~\cite{Bozek:2010vz,Xiao:2012uw,Jia:2014vja}.
The CME signal relative to the inclusive $\dgPP$ measurement is then given by~\cite{Xu:2017qfs}
\begin{equation}
    \begin{split}
		f_{\rm CME} & = \frac{\dg_{\rm CME}\{\psiPP\}}{\dgPP} =  \frac{A/a-1}{1/a^{2}-1}\,,
    \end{split}
	\label{eqD}
\end{equation}
where
\begin{equation}
	A=\dgSP/\dgPP\,.  
	\label{eqE}
\end{equation}
The above formalism applies even when the magnetic field direction does not coincide with $\psiSP$ as long as its fluctuations are independent from those of the $\psiPP$~\cite{Xu:2017qfs}. 
It is possible, however, that the magnetic field projection factor is not strictly $a$ because of final-state evolution effects on the charge separation~\cite{Shi:2017cpu}. A full study of this would require rigorous theoretical input and is beyond the scope of the present work.
There can be magnetic field contributions from participants; their contribution to $\dg$ 
follows the same projection as the background and is thus absorbed as part of the background.

{\em Data Analysis.} The data reported here are from Au+Au collisions taken by the STAR experiment 
at a nucleon-nucleon center-of-mass energy of $\snn = 200$~GeV 
in the years 2011, 2014 and 2016. 
A minimum-bias (MB) trigger was provided by a coincidence signal between the vertex position detectors located at forward/backward pseudo-rapidities ($\eta$) of $4.24<|\eta|<5.1$. 
Two zero-degree hadron calorimeters (ZDCs)~\cite{Adler:2000bd} 
cover $|\eta|>6.3$ and
intercept
spectator neutrons from the colliding beams. 
Shower maximum detectors (SMD) installed within the ZDCs measure
the positions of neutron-induced showers in the transverse plane~\cite{Adams:2005ca}.

The details of the STAR detector 
are described elsewhere~\cite{Ackermann:2002ad}. 
The main tracking device 
is the cylindrical time projection chamber (TPC)~\cite{Anderson:2003ur,Ackermann:1999kc}, providing 
full azimuthal coverage ($0<\phi<2\pi$) and an $\eta$ coverage of $-1.2<\eta<1.2$. 
Track trajectories are reconstructed from 3-dimensional hit points recorded by the TPC;
for a valid track, we require the number of hits ($N_{\rm hits}$) used in track fitting to be at least 10
out of a possible maximum ($N_{\rm max}$) of 45 is required for a valid track.
The TPC resides in a uniform 0.5~T magnetic field along the $-z$ direction, allowing determination of particle momenta from the track curvature for transverse momenta $\pt>0.15$~\GeVc.
The primary vertex of a collision is reconstructed from charged particle tracks. 
Events with primary vertices within 30~cm (year 2011) or 6~cm (years 2014 and 2016, taken with the heavy flavor tracker~\cite{Contin:2017mck}) longitudinally and 2~cm transversely from the geometrical center of the TPC are used, providing a total of 2.4 billion MB events. 
Events are also analyzed separately for positive and negative vertex $z$ samples to assess systematics from acceptance effects. 
Collision centrality is determined from the multiplicity of charged particles reconstructed in the TPC within a distance of closest approach (DCA) to the primary vertex of less than 3 cm 
and within an $\eta$ range of $|\eta|<0.5$~\cite{Abelev:2008ab}. 

Tracks used for the correlation analysis reported in this paper are required to have 
$N_{\rm hits}$ of at least 20 and DCA less than 1~cm. 
$N_{\rm hits}$ is varied to 15 and 25, and DCA is varied to 3.0, 2.0, and 0.8 cm to assess systematic uncertainties.
The fraction $N_{\rm hits}/N_{\rm max}$ is required to be greater than 0.52 to avoid double counting of split tracks.

Experimentally, the $\psiSP$ can be assessed by the first-order harmonic plane of spectator neutrons measured by the ZDC-SMD, and the $\psiPP$ by the second-order harmonic plane of mid-rapidity particles measured by the TPC~\cite{Poskanzer:1998yz}. 
In the rest of the paper, we refer to the former as $\psiZDC$ and the latter as $\psiTPC$. 
In this analysis, the $\gamma$ and $v_{2}$ are calculated by
$\gamma = \langle \cos(\phi_{\alpha}+\phi_{\beta}-2\psirec)\rangle / R\,$ 
and
$v_{2} = \langle \cos 2(\phi_{\alpha, \beta}-\psirec)\rangle / R\,,$
where $\psirec$ is either $\psiZDC$ or $\psiTPC$,
and $R$ is the corresponding resolution~\cite{Poskanzer:1998yz}.
For $\psiTPC$, a $\phi$-dependent weight is applied to account for track detection efficiency, and the $R$ is calculated from the correlations between two TPC sub-events (see below)~\cite{Poskanzer:1998yz}.
For $\psiZDC$, an event-plane vector is determined from the measured energy distribution combining both ZDCs, and the $R$ is calculated from the correlations between their event-plane vectors~\cite{Poskanzer:1998yz,Adams:2005ca}. 
The standard recentering and shifting techniques~\cite{Poskanzer:1998yz} are applied. 

The same particles of interest (POI), denoted by $\alpha$ and $\beta$, are used for $\gamma$ and $v_{2}$ with $\pT$ from 0.2 to 2~\GeVc. 
The $\phi$-dependent track efficiency is corrected for the POIs.
A $\pT$-dependent efficiency correction 
does not reveal any systematic effect.
Two methods are employed in this analysis. 
The first one, referred to as the ``full-event" method, uses particles from $|\eta|<1$ as the POIs. 
A third particle $(c)$ from the same acceptance is used in place of $\psiTPC$, and $R$ equals the particle $v_{2,c}$~\cite{Abelev:2009ad}. 
For this method, another $\pT$ range from 0.2 to 1~\GeVc\ is also analyzed for the POIs 
to explore possible $\pT$ dependence of the CME signal, speculated to be dominant at low $\pT$~\cite{Kharzeev:2007jp}. 
The second method, referred to as the ``sub-event" method, divides the TPC particles into two sub-events symmetric about mid-rapidity~\cite{Poskanzer:1998yz}, $\Delta\eta_{\rm sub}/2<|\eta|<1$ with an $\eta$ gap ($\Delta\eta_{\rm sub}$) in-between, where the POIs are from one sub-event and the $\psiTPC$ is reconstructed from the other. 
This procedure reduces non-flow correlations that are short-ranged, such as those due to resonance decays and jets~\cite{Wang:2009kd,Petersen:2010di,Zhao:2019kyk}.
We perform the analyses with $\Delta\eta_{\rm sub}=0.1$ and $0.3$.

To assess systematic uncertainties,
the full analysis is repeated for each cut variation and results from different years are combined at the end. 
Data from the various centralities are combined and compared to the default case
In this way, the (anti-)correlations in the uncertainties are properly taken into account.
The influence of statistical uncertainties in systematic error estimation is treated as in Ref.~\cite{Barlow:2002yb}.
For each source when multiple variations are used, the systematic uncertainty is taken as the RMS. 
In order to minimize fluctuations due to the limited statistics, the systematic uncertainty for the entire 20--80\% centrality range is also evaluated. The larger value between it and the 20--50\% (or 50--80\%) range is quoted, 
unless both are zero (i.e., results are consistent within statistical fluctuations); in this case, the systematic uncertainties evaluated from the individual centralities are presented.

For the 20--50\% centrality, the absolute systematic uncertainties on $\mean{f_{\rm CME}}$ for $0.2 < p_{T} < 2.0$  \GeVc\ with the full-event method are 2.2$\%$ and 1.3$\%$ for the number of hits and DCA variations, respectively.
The $\mean{f_{\rm CME}}$ results from positive and negative vertex $z$ events are consistent within statistical uncertainties for the combined 20--50\% centrality; therefore systematic uncertainties evaluated for individual centralities are presented. 
For the 50--80\% centrality range, the combined systematic uncertainty is used.
The variations in the results among the three run periods 
beyond statistical fluctuations are taken as part of the systematic uncertainties. 
For $\mean{f_{\rm CME}}$, the results are consistent within statistical uncertainties.
To investigate the effect of $\psiZDC$ determination, analyses are also performed using only a single ZDC side for $\psiZDC$ 
as well as an arithmetic average of the $\psiZDC$ values from the two sides.
The results are consistent with the default case within statistical uncertainties.
The systematic uncertainties from the various sources are added in quadrature 
and are quoted for one standard deviation.

\begin{figure*}[hbt]
	\begin{center}
		\includegraphics[width=0.32\textwidth]{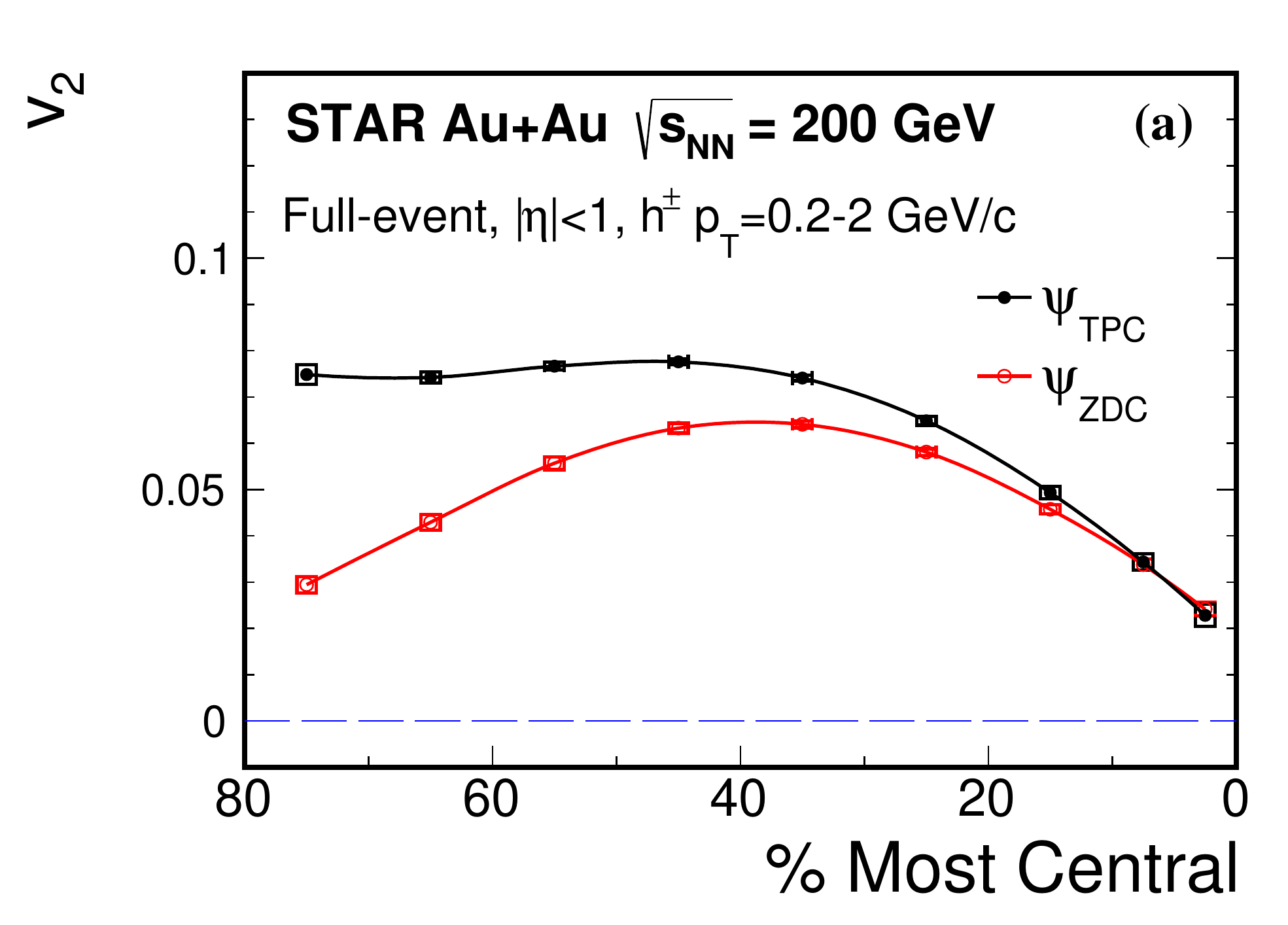}
		\includegraphics[width=0.32\textwidth]{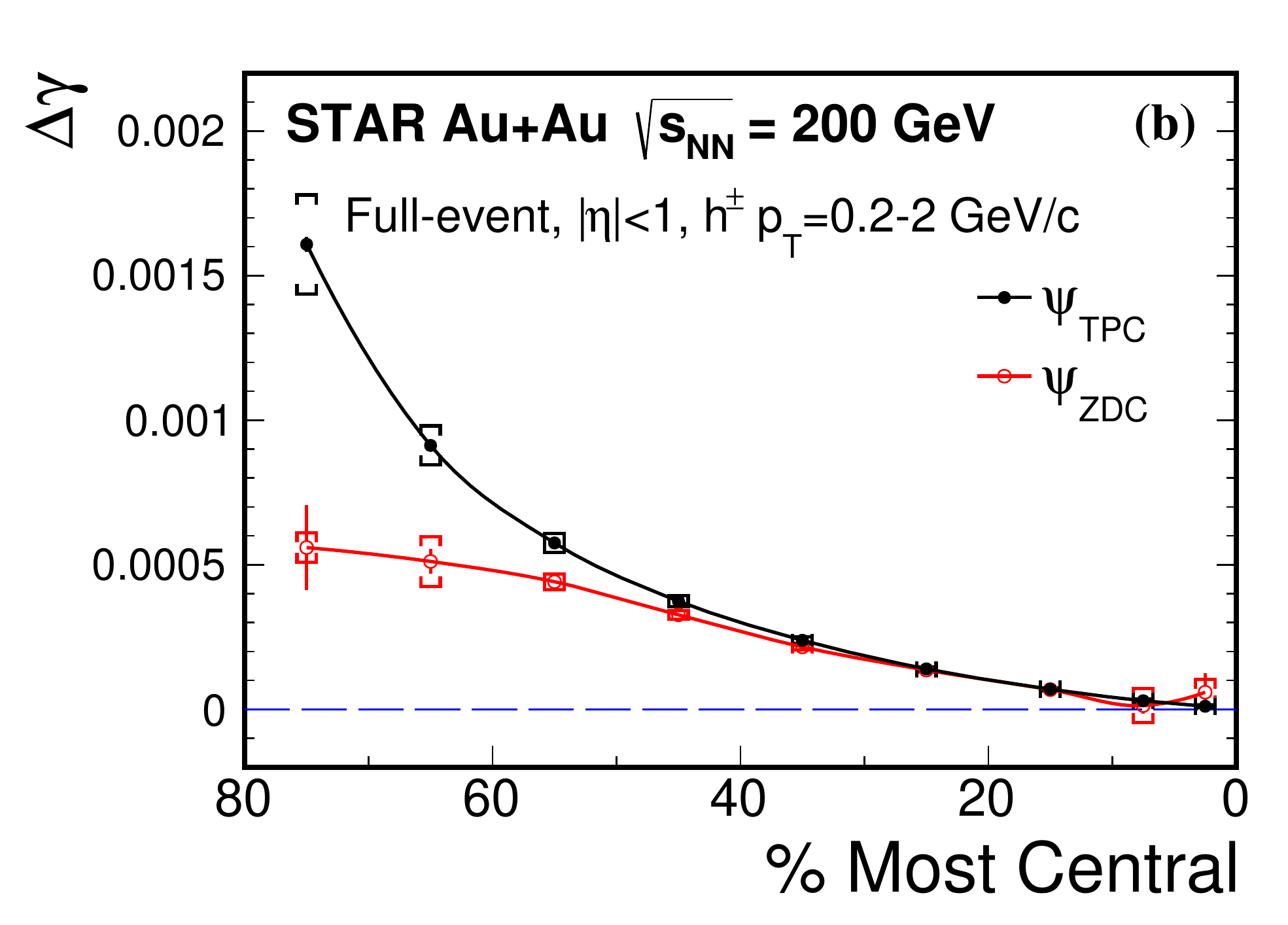}
		\includegraphics[width=0.32\textwidth]{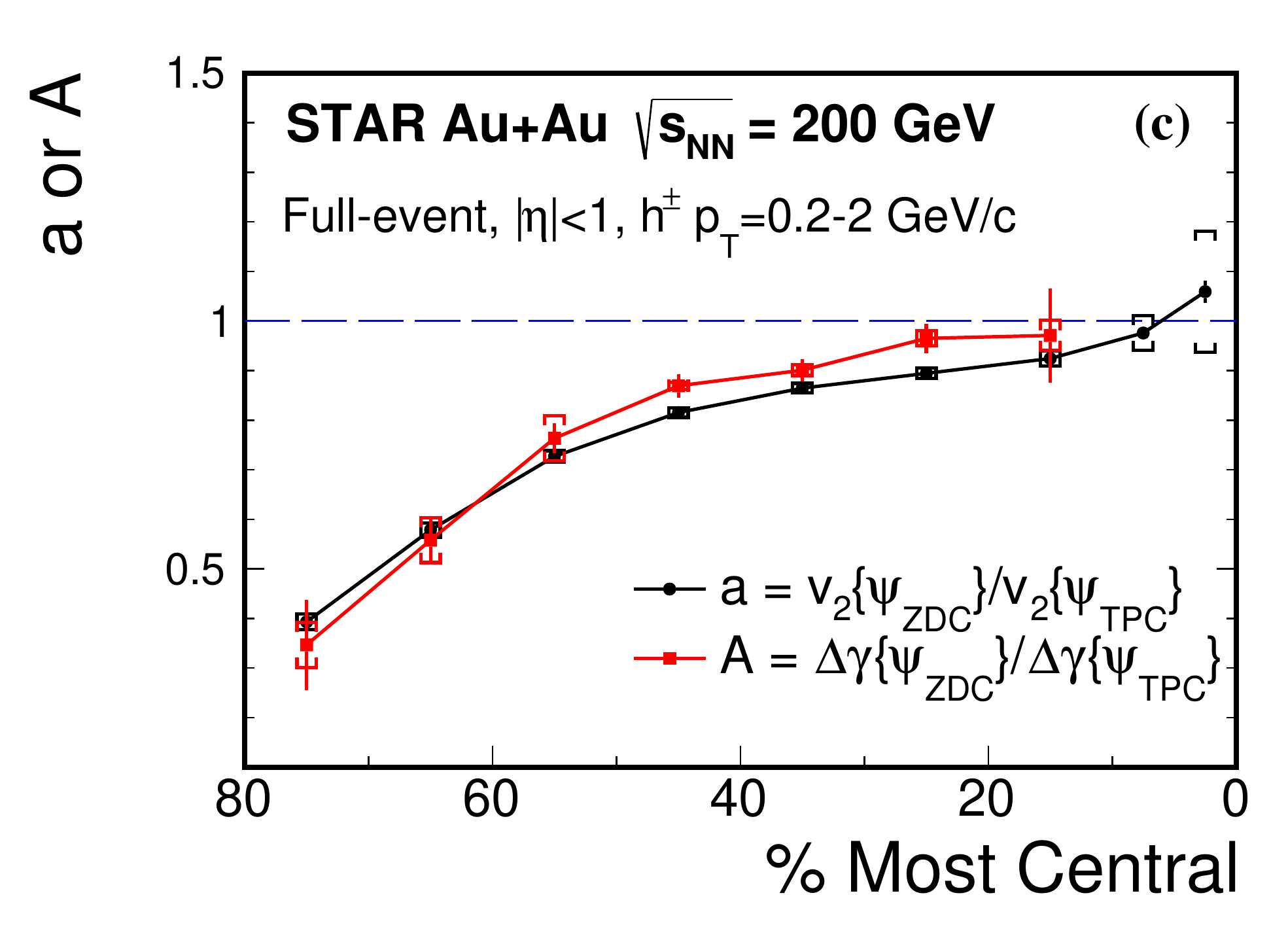}
		\caption{
		The centrality dependencies of the $v_{2}$ (a) and $\dg$ (b)
		measured with respect to $\psiZDC$ and $\psiTPC$ from the full-event method. 
		Panel (c) presents the ratios $a$ and $A$.
		Error bars show statistical uncertainties; the caps indicate the systematic uncertainties.}
		\label{fig1}
	\end{center}
\end{figure*}

{\em Results and discussions.} Figure~\ref{fig1}, panels (a) and (b) show, respectively, the measured $v_{2}$ and $\dg$ with respect to the $\psiZDC$ and $\psiTPC$ from the full-event method with $0.2<\pT<2$~\gevc\ in Au+Au collisions at $\snn=200$~GeV as a function of centrality. 
The $v_2\{\psiZDC\}$ is smaller than $v_2\{\psiTPC\}$, as expected; 
the $\dg\{\psiZDC\}$ is also smaller than $\dg\{\psiTPC\}$, as expected if they are dominated by $v_2$ backgrounds.
Figure~\ref{fig1}(c) shows the quantites $a=v_2\{\psiZDC\}/v_2\{\psiTPC\}$ and $A=\dg\{\psiZDC\}/\dg\{\psiTPC\}$ as functions of centrality.
Their values are found to be nearly identical over the full centrality range, indicating the dominance of background contributions in $\dg$.

\begin{figure*}[hbt]
	\centering
	\includegraphics[width=0.32\textwidth]{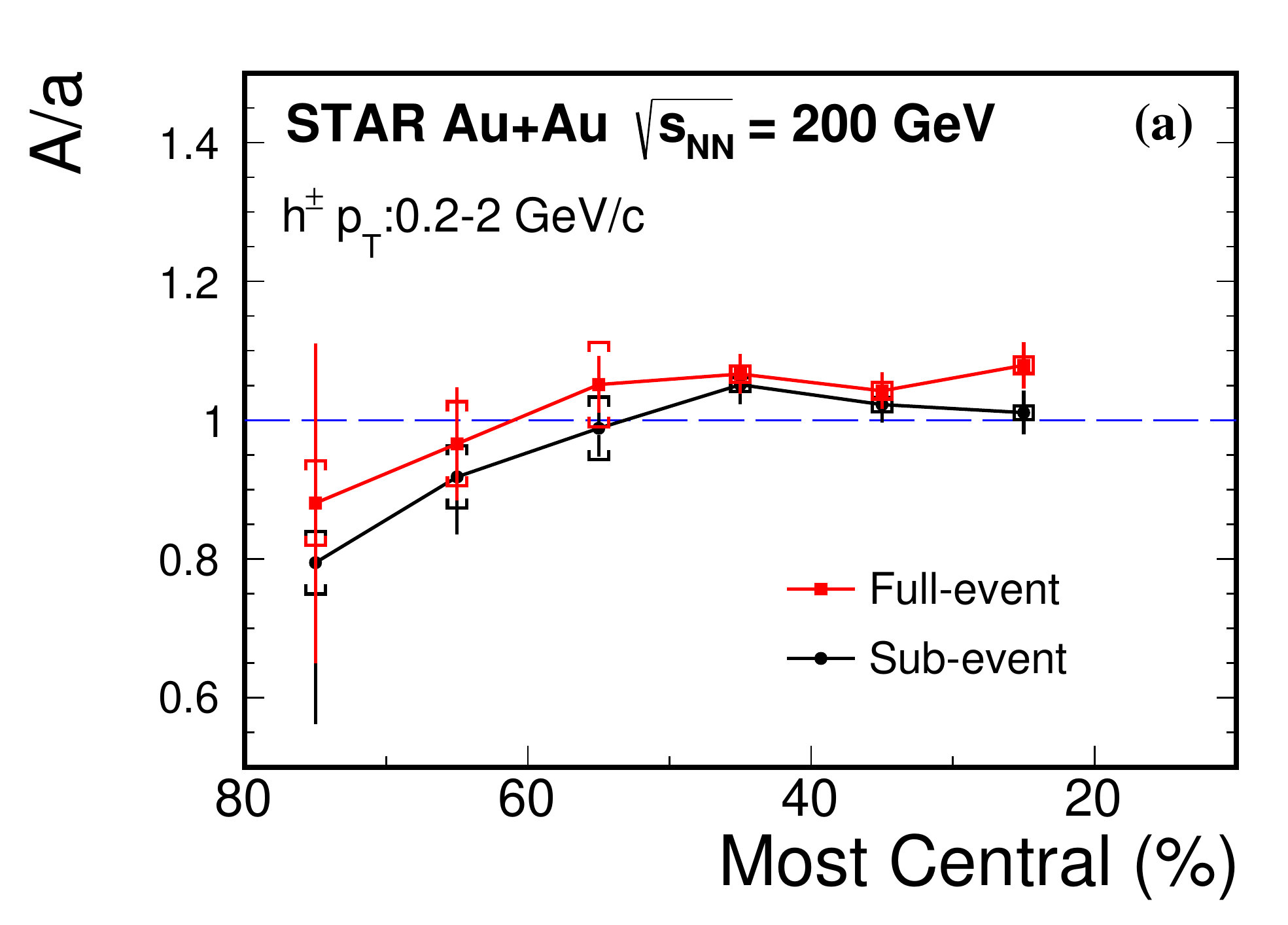}
	\includegraphics[width=0.32\textwidth]{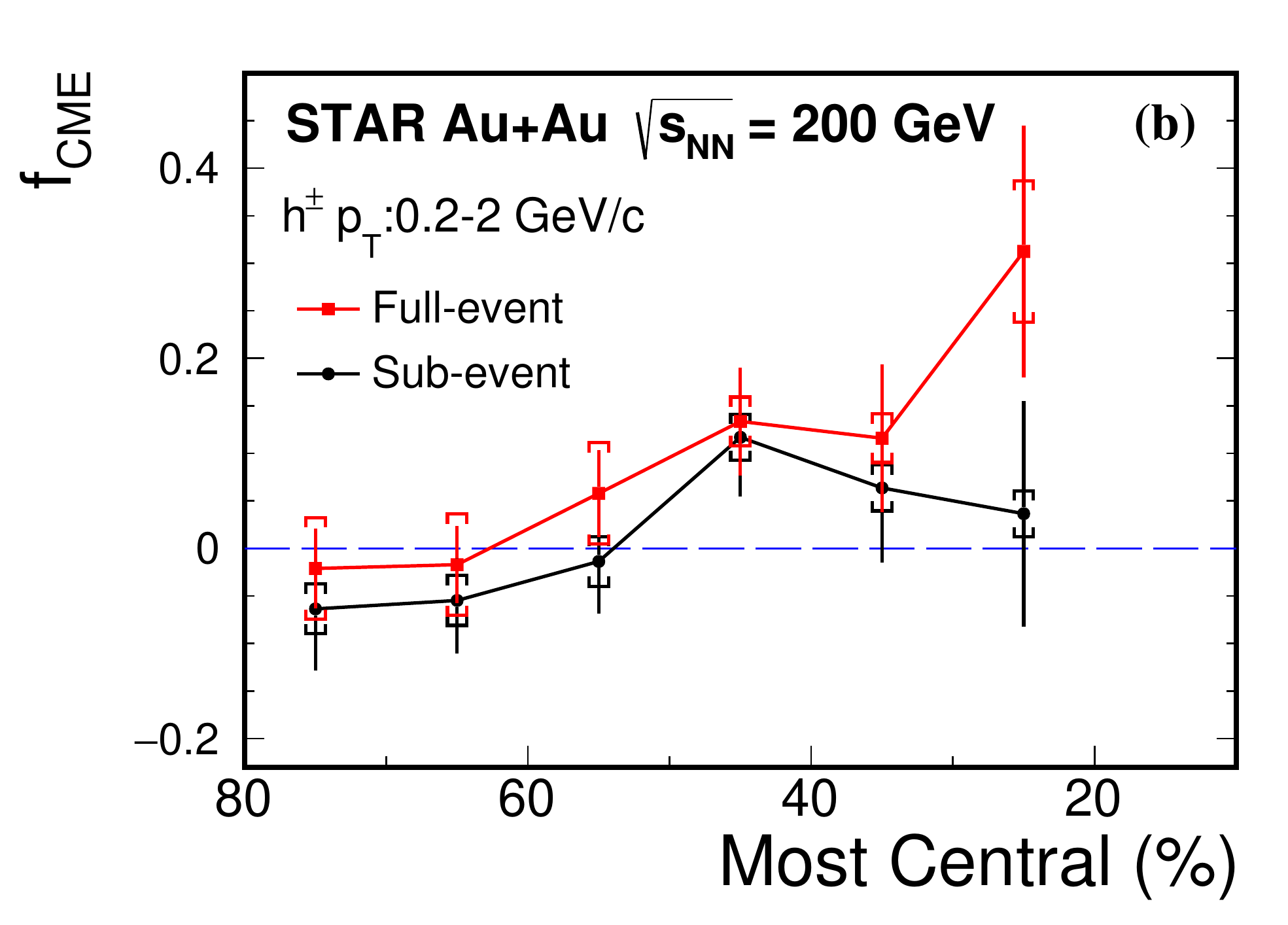}
	\includegraphics[width=0.32\textwidth]{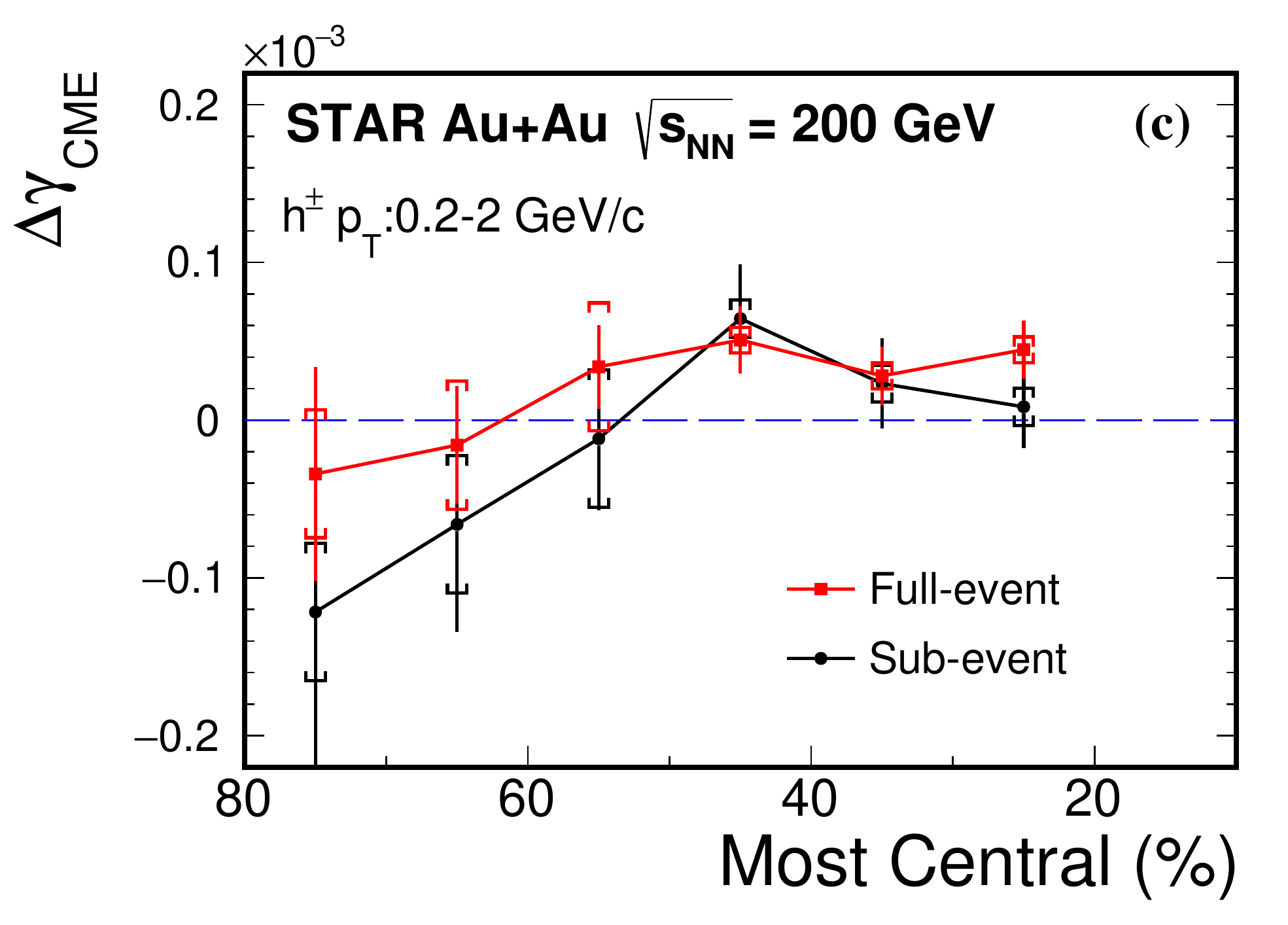}
	\caption{
		The $A/a$ ratio (a), the extracted $f_{\rm CME}$ (b), and $\dg_{\rm CME}$ (c) as functions of the collision centrality from the full-event and sub-event ($\Delta\eta_{\rm sub}=0.1$) methods.
		Error bars show statistical uncertainties; the caps indicate the systematic uncertainties.
		}
	\label{fig2}
\end{figure*}

Figure~\ref{fig2}(a) shows the $A/a$ ratio from both the full-event and sub-event methods, for $0.2<\pT<2$~\gevc.  
A value of $A/a>1$ would indicate the possible existence of a CME signal.
Figure~\ref{fig2}(b) shows the centrality dependence of $f_{\rm CME}$, the possible CME signal relative to the inclusive measurement $\dg\{\psiTPC\}$, extracted by Eq.~(\ref{eqD}).
Figure~\ref{fig2}(c) shows the absolute magnitude of the signal, 
$\dg_{\rm CME}\equiv\dg_{\rm CME}\{\psiTPC\}=f_{\rm CME}\dg\{\psiTPC\}$, 
as a function of centrality. 

\begin{figure*}[hbt]
	\centering
	\includegraphics[width=0.49\textwidth]{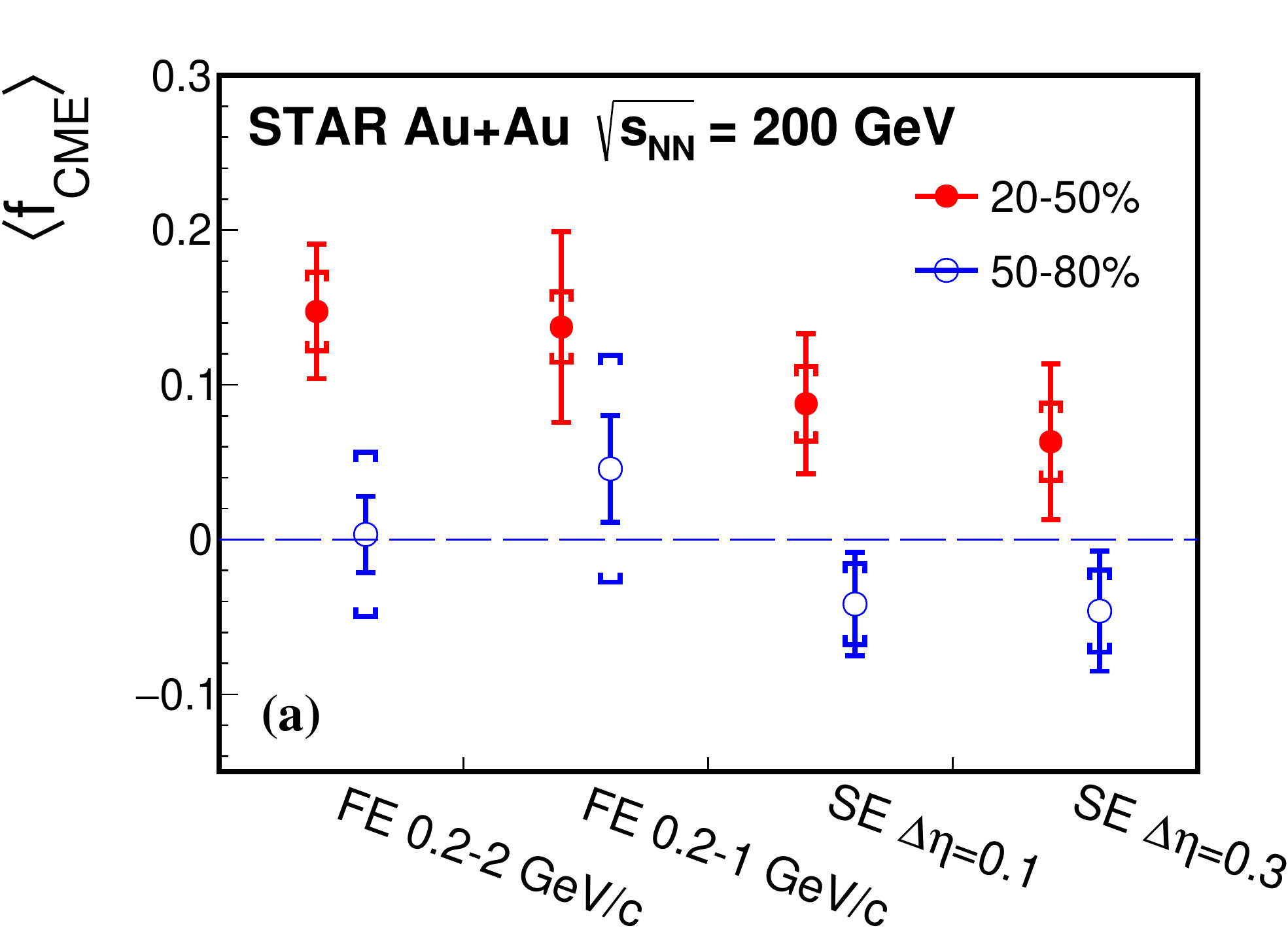}
	\includegraphics[width=0.49\textwidth]{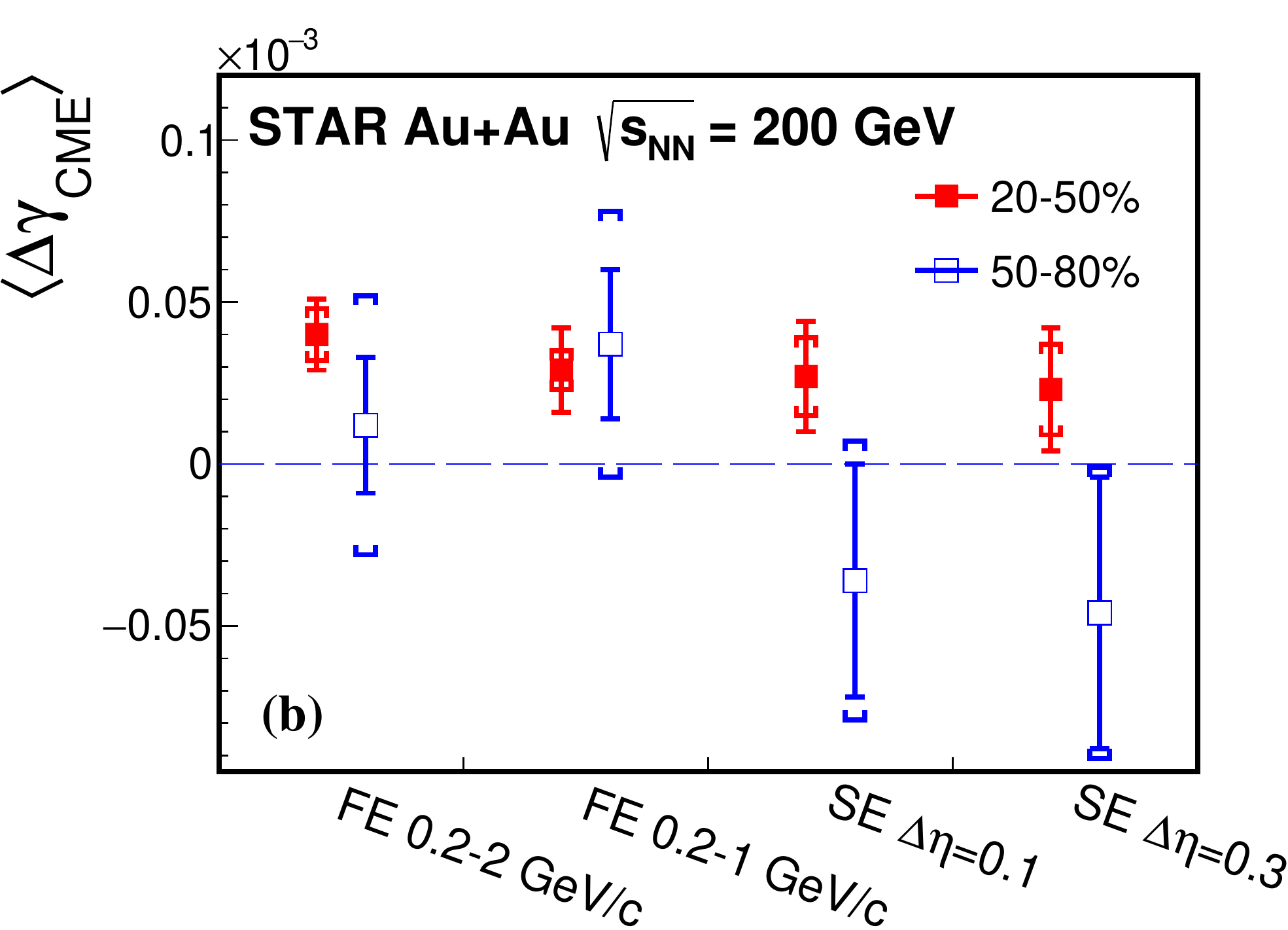}
	\caption{
		The flow-background removed $\mean{f_{\rm CME}}$ (a) and $\mean{\dg_{\rm CME}}$ (b) signal in 50--80\% (open markers) and 20--50\% (solid markers) centrality Au+Au collisions at \sNN = 200 GeV, extracted by various analysis methods (FE: full-event, SE: sub-event) and kinematic cuts.
		Error bars show statistical uncertainties; the caps indicate the systematic uncertainties.
		}
	\label{CMEAuAu}
\end{figure*}

Table~\ref{table1} reports $\mean{f_{\rm CME}}$ and $\mean{\dg_{\rm CME}}$, averaged over 20--50\% and 50--80\% centrality ranges, along with the inclusive $\mean{\dg\{\psiTPC\}}$. 
Both the full-event and sub-event methods are tabulated. 
The results are shown in Fig.~\ref{CMEAuAu}, and are consistent with zero in the 50--80\% peripheral centrality range. For the 20--50\% centrality range, hint of the signal deviating from zero is seen with 1--3 standard deviations, depending on the analysis method. 
Note that the statistical and systematic uncertainties are not completely independent among the data points because the same overall data sample is used in the various methods.

Since the CME is speculated to be a low-$\pT$ phenomenon~\cite{Kharzeev:2007jp}, we have analyzed a lower $\pT$ range $0.2<\pT<1$ \GeVc\ for the POI for the full-event method, as shown in Fig.~\ref{CMEAuAu}.
Given the large uncertainties 
we cannot draw conclusions concerning the relative magnitude of $f_{\rm CME}$ or $\dg_{\rm CME}$ between the two $\pT$ ranges.

\begin{table*}[htpb]
\centering
\caption{
	The inclusive $\mean{\dg\{\psiTPC\}}$ and the extracted $\mean{f_{\rm CME}}$ and $\mean{\dg_{\rm CME}}$, averaged over 20--50$\%$ and 50--80\% centrality ranges in Au+Au collisions at \sNN = 200 GeV from the full-event method (with two POI $\pT$ ranges) and the sub-event method (with two $\eta$ gaps). The first quoted uncertainty is statistical and the second systematic.
	}
\begin{tabular}{llrrr}
\hline
Centrality\hspace{5mm} & Method\hspace{3cm} &
\hspace{10mm}$\mean{\dg_{\rm inc}}$ ($\times10^{-4}$) &
\hspace{14mm}$\mean{f_{\rm CME}}$ (\%) &
\hspace{8mm}$\mean{\dg_{\rm CME}}$ ($\times10^{-4}$)\\ 
\hline
\multirow{4}{*}{20--50\%}
& full-event, $p_{T}$=0.2--2 \GeVc
& $1.89\pm0.01\pm0.10$ & $14.7\pm4.3\pm2.6$ & $ 0.40\pm0.11\pm0.08$ \\
& full-event, $p_{T}$=0.2--1 \GeVc
& $1.48\pm0.01\pm0.07$ & $13.7\pm6.2\pm2.3$ & $ 0.29\pm0.13\pm0.06$ \\
& sub-event, $\Delta\eta_{\rm sub}$=0.1, $p_{T}$=0.2--2 \GeVc
& $2.84\pm0.01\pm0.15$ & $ 8.8\pm4.5\pm2.4$ & $ 0.27\pm0.17\pm0.12$ \\
& sub-event, $\Delta\eta_{\rm sub}$=0.3, $p_{T}$=0.2--2 \GeVc
& $2.94\pm0.01\pm0.15$ & $ 6.3\pm5.0\pm2.5$ & $ 0.23\pm0.19\pm0.14$ \\
\hline
\multirow{4}{*}{50--80\%}
& full-event, $p_{T}$=0.2--2 \GeVc
& $6.31\pm0.03\pm0.38$ & $ 0.3\pm2.5\pm5.3$ & $ 0.12\pm0.21\pm0.40$ \\
& full-event, $p_{T}$=0.2--1 \GeVc
& $5.19\pm0.04\pm0.33$ & $ 4.6\pm3.4\pm7.3$ & $ 0.37\pm0.23\pm0.41$ \\
& sub-event, $\Delta\eta_{\rm sub}$=0.1, $p_{T}$=0.2--2 \GeVc
& $8.72\pm0.06\pm0.41$ & $-4.2\pm3.4\pm2.6$ & $-0.36\pm0.36\pm0.43$ \\
& sub-event, $\Delta\eta_{\rm sub}$=0.3, $p_{T}$=0.2--2 \GeVc
& $8.89\pm0.07\pm0.40$ & $-4.6\pm3.9\pm2.7$ & $-0.46\pm0.43\pm0.45$ \\
\hline
\end{tabular}
\label{table1}
\end{table*}

A key assumption made in this analysis is that the flow background is proportional to the final-state hadron $v_2$~\cite{Xu:2017qfs}. 
This assumption may not strictly hold because of the presence of non-flow. 
For example, two-particle correlations contribute positively to $v_2\{\psiTPC\}$, which would reduce $a$, yielding an increased $f_{\rm CME}$. 
Three-particle (e.g.~dijet) correlations 
could significantly increase $\dg\{\psiTPC\}$, 
which would reduce $A$, and thus cause a decreased $f_{\rm CME}$. 
The latter may have contributed to the negative $f_{\rm CME}$ in peripheral collisions (modulo large uncertainties)~\cite{Feng:2021pgf}.
The relative strengths of those effects are unknown a priori. 
The measured $f_{\rm CME}$ and $\dg_{\rm CME}$ can, therefore, still be contaminated by non-flow effects. 
In order to mitigate non-flow effects, we have analyzed data using the sub-event method with two $\eta$ gaps, as also shown in Fig.~\ref{CMEAuAu}. 
The extracted $f_{\rm CME}$ and $\dg_{\rm CME}$ are of reduced significance because of the smaller particle pair statistics
with the sub-event method.
It is noteworthy that our result is consistent, within one standard deviation, with the previously extracted $f_{\rm CME}$=$(2 \pm 4 \pm 5)\%$~\cite{Adam:2020zsu} (also from the sub-event method) exploiting the pair invariant mass~\cite{Zhao:2017nfq}. 
The method exploited in the present work uses additional information from the ZDCs taking advantage of the PP and SP fluctuations.

Recently STAR has released results from a blind analysis of isobar collisions~\cite{STAR:2021mii},
which offer improved discrimination between the possible CME signal and the known backgrounds.
A significance of 3 standard deviations is expected if the CME fraction is 10\% in isobar collisions~\cite{STAR:1718bur,Deng:2016knn}. 
However, no evidence of the CME has been observed, suggesting that the CME fraction in isobar collisions is significantly smaller than 10\%. This would be consistent with the data reported here if the CME signal to background ratio is substantially reduced from Au+Au to isobar collisions as suggested in Ref.~\cite{Feng:2021oub}.

{\em Conclusions.}
In summary, we have reported measurements of the elliptic flow anisotropy $v_2$ and three-particle correlator $\dg$ with respect to 
the first-order harmonic plane from the zero-degree calorimeters, 
$\psiZDC$, and the second-order harmonic plane from the time projection chamber, $\psiTPC$. 
We used the full-event method where the particles of interest POI and $\psiTPC$ are both from the $|\eta|<1$ range,
and studied two $\pT$ ranges for the POI. 
We also used the sub-event method where the POI and $\psiTPC$ are from two sub-events, and we applied two $\eta$ gaps between the sub-events. 
The inclusive $\dg$ measurements with respect to $\psiZDC$ and $\psiTPC$ are found to be largely dominated by backgrounds, consistent with conclusions from previous measurements. 
Because $\psiZDC$ aligns better with the spectator proton plane and $\psiTPC$ aligns better with the $v_2$ harmonic plane, 
these measurements can be used to extract the possible CME signals, assuming that the background is proportional to $v_2$ and the magnetic field is determined by the spectator protons.
Under these assumptions, the possible CME signals are extracted using the new method in this paper. 
Some indication of finite signals is seen in 20--50\% Au+Au collisions. 
However, non-flow effects (especially for the full-event method without $\eta$ gap) may still be present that warrant further investigation.

{\em Acknowledgments.} We thank the RHIC Operations Group and RCF at BNL, the NERSC Center at LBNL, and the Open Science Grid consortium for providing resources and support.  This work was supported in part by the Office of Nuclear Physics within the U.S. DOE Office of Science, the U.S. National Science Foundation, the Ministry of Education and Science of the Russian Federation, National Natural Science Foundation of China, Chinese Academy of Science, the Ministry of Science and Technology of China and the Chinese Ministry of Education, the Higher Education Sprout Project by Ministry of Education at NCKU, the National Research Foundation of Korea, Czech Science Foundation and Ministry of Education, Youth and Sports of the Czech Republic, Hungarian National Research, Development and Innovation Office, New National Excellency Programme of the Hungarian Ministry of Human Capacities, Department of Atomic Energy and Department of Science and Technology of the Government of India, the National Science Centre of Poland, the Ministry  of Science, Education and Sports of the Republic of Croatia, RosAtom of Russia and German Bundesministerium fur Bildung, Wissenschaft, Forschung and Technologie (BMBF), Helmholtz Association, Ministry of Education, Culture, Sports, Science, and Technology (MEXT) and Japan Society for the Promotion of Science (JSPS).


\bibliography{ref}

\end{document}